\documentclass[twocolumn,showpacs,preprintnumbers,amsmath,amssymb,prl]{revtex4}
 \usepackage[dvips]{graphicx}
 \usepackage{dcolumn}
 \usepackage{bm}

\begin{document}

\title{Tetrahedral clustering in molten lithium under pressure}

\author{Isaac Tamblyn$^1$, Jean-Yves Raty$^2$, and Stanimir A. Bonev$^1$}
\email[Electronic address: ]{stanimir.bonev@dal.ca}
\affiliation{$^1$ Department of Physics, Dalhousie University, Halifax, NS, B3H 3J5, Canada \\ $^2$ FNRS-University of Liege, Sart-Tilman 4000, Belgium }

\date{\today}

\begin{abstract}

A series of electronic and structural transitions are predicted in molten lithium from first principles. A new phase with tetrahedral local order characteristic of $sp^3$ bonded materials and poor electrical conductivity is found at pressures above 150~GPa and temperatures as high as 1000~K. Despite the lack of covalent bonding, weakly bound tetrahedral clusters with finite lifetimes are predicted to exist. The stabilization of this phase in lithium involves a unique mechanism of strong electron localization in interstitial regions and interactions among core electrons. The calculations provide evidence for anomalous melting above 20 GPa, with a melting temperature decreasing below 300 K, and point towards the existence of novel low-symmetry crystalline phases.

\end{abstract}


\maketitle

%
%

The properties of materials under high pressure can change in profound and unexpected ways. At ambient pressure ($P$) and temperature ($T$), Li can be regarded as the prototype for an ideal metal. However, theoretical studies \cite{1} have predicted that as a result of increased core-valence electron interactions, low-$T$ dense solid Li would undergo a series of symmetry-breaking transitions, culminating in a Li$_2$-paired crystal with semi-metallic properties at pressures above 150 GPa. Besides the regular transition at 7.5 GPa from body-centered cubic ($bcc$) to face-centered cubic ($fcc$) structure \cite{2}, measurements to date \cite{3} have confirmed only an initial transition at 39 GPa from fcc to a less compact structure with a 16-atom cubic cell ($cI16$ space group) via an intermediate disordered phase ($hR1$). The existence and properties of even lower-coordinated structures at higher pressures remains an open question due to the lack of experimental data and difficulties associated with conclusive theoretical predictions of low-temperature thermodynamic crystalline stability. 

It was shown recently \cite{4} that solid-solid transitions in Na are preceded by analogous changes in its liquid but at much lower pressure. Similar behavior was also suggested for other light alkalis.
This raises the intriguing question, answered in what follows, of whether the existence of low-symmetry structures in Li \cite{1,3,6,7,8} could be established by investigating its molten phase. Molecular dynamics simulations of liquids have the advantage of depending minimally on initial conditions and are not subject to the inherent bias of comparing a finite number of solid configurations. 
Finding low-coordinated structures in molecular dynamics simulations of a liquid, in addition to being remarkable in itself, would therefore constitute rather conclusive evidence for the existence of similar crystalline phases. 

%
%

We have carried out first principles molecular dynamics (FPMD) 
simulation of liquid Li in the density range corresponding 
to $3.06 > r_s > 1.60$ (where $\frac{4}{3}\pi(r_{s}a_{0})^{3} = V/N$, $a_{0}$ 
is the Bohr radius, $V$ the volume and $N$ the number of valence electrons) at temperatures up to 3000 K, and of the known crystalline phases: $bcc$, $fcc$, and $cI16$, between 0 and 90 GPa. Density functional theory with a plane wave basis set and the Born-Oppenheimer approximation were used for all simulations \cite{dft}. They were performed in the $NVT$ ensemble with cubic supercells and periodic boundary conditions. For the simulations of $bcc$ and $cI16$ solids, and all liquids we used 128 atom supercells with $\Gamma$-point sampling of the Brillouin zone.
Simulations of the $fcc$ solids were performed with 108 atoms and 8 ${\mathbf k}$-points (the $\Gamma$-point only was found to be insufficient). Most simulations ran for at least 10 ps, with some up to 20 ps, while for obtaining the melting temperatures we have computations as long as 200 ps. Extensive convergence tests were carried out for size effects, various simulation parameters, and the validity of the pseudopotential approximation \cite{convergence}. Good agreement was also obtained with existing liquid experimental data \cite{26}.

%
%

\begin{figure}[tbh]
  \hspace*{-5mm}
  \includegraphics[height=0.25\textheight,clip]{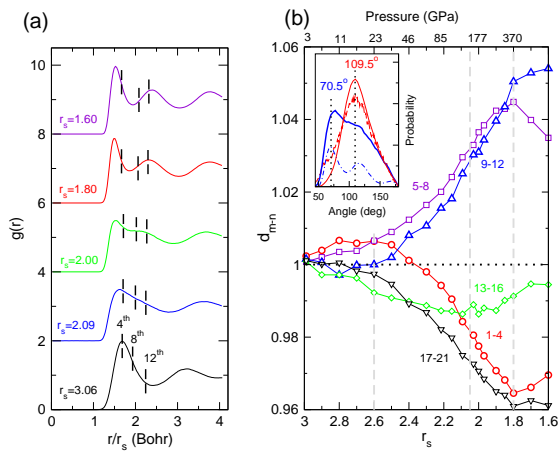}
  \caption{\label{fig_str} Structural changes in molten Li
along the 1000~K isotherm. (a) Pair correlation function, $g(r)$, for selected densities. Vertical dashed lines indicate the average positions of the 4$^{\mathrm{th}}$, 8$^{\mathrm{th}}$, and 12$^{\mathrm{th}}$ neighbors. (b) Evolution of density-rescaled average interatomic distances with pressure and density. First, average interatomic distances are computed to the first ($r_1$), second ($r_2$), and so on, neighbors. Next, density-rescaled distances are calculated as $d_i = (r_i/r_s)/(r_i^0/r_s^0)$, where $r_s^0=3.06$ and $r_i^0$ refers to distances obtained at $r_s^0$. Finally, $d_{m-n}$ is the averaged value of $d_i$ for neighbors, $i$, from $m$ to $n$. The corresponding values of m and n are indicated next to each curve. The inset shows bond angle distributions between the first two neighbors in molten Li (solid curves) for $r_s = 3.06$ (blue) and 1.80 (red), of solid bcc Li (dashed blue line) at 500 K and $r_s = 3.06$, and of molten diamond \cite{28} around 250 GPa (dashed red curve). Angles among nearest neighbors in ideal bcc (70.50$^o$) and diamond (109.50$^o$) crystals are indicated. The bcc data are scaled by a factor of 1/3 relative to the other angle distributions.}
\end{figure}

Results for the structural analysis of the liquid along a 1000~K isotherm are presented in Fig.~\ref{fig_str}. Initially, the first peak of the  pair correlation function, $g(r)$, broadens (Fig.~\ref{fig_str}(a)) in a way previously observed in Na \cite{4}. Upon further compression, it splits entirely, indicating significant further lowering in the coordination. For a more detailed analysis of the structure, we examine the evolution of neighbor distances with increasing density (Fig.~\ref{fig_str}(b)). While there is no indication for discontinuous liquid-liquid transitions, several distinct regions with different liquid structures can be identified, which correlate well with the electronic and melting properties discussed further below. These are: (i) $r_s \gtrsim 2.60$ ($P < 23$~GPa); (ii) $2.60 \gtrsim r_s \gtrsim 2.05$ (23 GPa $< P < 150$~GPa); and (iii) $r_s \lesssim 2.05$ ($P > 150$ GPa). The initial changes are analogous to those observed in liquid Na \cite{4} - a transition from a $bcc$-like to an $fcc$-like local order in (i), followed in (ii) by lowering in the coordination (number of neighbors under the symmetrized first peak of $g(r)$) to $8+4+\ldots$, and the liquid acquiring a $cI16$-like local order.

In (iii),  the density-rescaled average distances to neighbors 1-4 continue to contract even faster, while the next eight, as well as 13-16, move away. As a result, by $r_s = 1.80$ ($P \sim 370$~GPa), the first coordination shell completely splits in two, the coordination becomes only four and remains roughly so up to at least $r_s = 1.6$ ($P \sim 810$~GPa). The distribution of angles among the nearest neighbors also becomes rather unexpected (inset in Fig.~\ref{fig_str}(b)) as it has a peak at 109.5$^o$. The parallel with the liquids of materials with sp$^3$ bonding is indeed striking when compared to the corresponding angle distribution in molten carbon, obtained by melting diamond at similar pressure (inset in Fig.~\ref{fig_str}(b)). Therefore, there is a large pressure range, 150 GPa $< P <$ 810 GPa, for which we predict that Li has tetrahedral local order, hitherto not seen in a liquid metal, but characteristic of semiconductor liquids with $sp^3$ bonding.

\begin{figure}[tbh]
  \hspace*{-5mm}
  \includegraphics[height=0.22\textheight,clip]{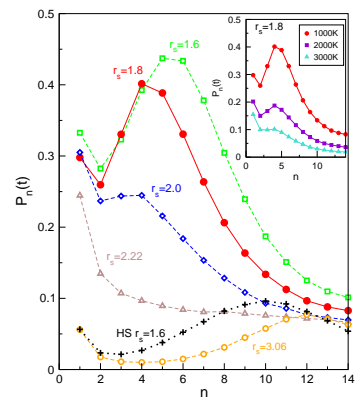}
  \caption{\label{fig_lifetimes} Survival probability, $P_n(t)$, of clusters as a function of their size, $n$. Results are shown for Li
and for a hard sphere (HS) liquid.
The times at which the $P_n(t)$'s are evaluated are: 43 fs (for $r_s=3.06)$, 27 fs ($r_s=2.22$), 20 fs ($r_s=2.00$), 15 fs ($r_s=1.80$) and 12 fs ($r_s=1.60$) for Li, and 20 fs for the HS liquid. The HS radius is chosen such that the self-diffusion constant and the short-distance $g(r)$'s of the HS and Li liquids are roughly the same. Computations with HS's at lower densities show rapidly decreasing values of $P_n(t)$; already an order of magnitude lower at $r_s=2.0$. The inset shows survival probability of clusters as a function of temperature at $r_s=1.80$.}
\end{figure}
%
%

The apparent structural similarity with covalently bonded liquids has prompted us to investigate the persistence probability of Li clusters. For this purpose, we evaluate a function $P_n(t)$, which gives the statistical probability for the first $(n-1)$ nearest neighbours found around an atom at time $t$ to be the same as those found at time $t=0$. For completely uncorrelated particles, $P_n(t)$ would drop monotonically with increasing $n$ for any fixed $t$. This is a result of the fact that the larger the cluster the more particles there are that can leave the cluster during any time interval. For the same reason, $P_n(t)$ should be a decreasing function of $n$ for a sufficiently large $t$ even if there is some weak binding among the atoms. For example, if $p$ is the average probability for a given particle to leave a cluster during a time interval $t$, then the survivor probability of that cluster is $P_n(t) \sim (1-p)(n-1)$ for large $t$ (and neglecting particles re-entering the cluster).

In order to study cases with weak metastability, the time chosen for the evaluation of $P_n(t)$ should be sufficiently short such that results are non-trivial. The values of $t$ at which we have evaluated $P_n(t)$ for each density are listed in the caption of Fig.~\ref{fig_lifetimes}. The ratios among them are roughly equal to the ratios of the periods of atomic vibrations at different densities, which have been estimated from a velocity autocorrelation analysis. This allows us to quantitatively compare the metastability of clusters for a range of densities, over which diffusion rates and vibrational properties differ significantly.
As shown in Fig.~\ref{fig_lifetimes}, $P_n(t)$ changes qualitatively when density is increased. The appearance of a peak at $n=5$ is evidence for a weak metastability of the tetrahedral clusters. 
To estimate the importance of caging effects \cite{9}, which could appear in a compressed liquid, we have also carried out simulations of hard-sphere liquids. The comparison with Li (Fig.~\ref{fig_lifetimes}) demonstrates that the peak in $P_n(t)$ of Li is much higher, indicating significantly stronger interatomic correlations.

\begin{figure}[tb!]
  \hspace*{-5mm}
  \includegraphics[height=0.22\textheight,clip]{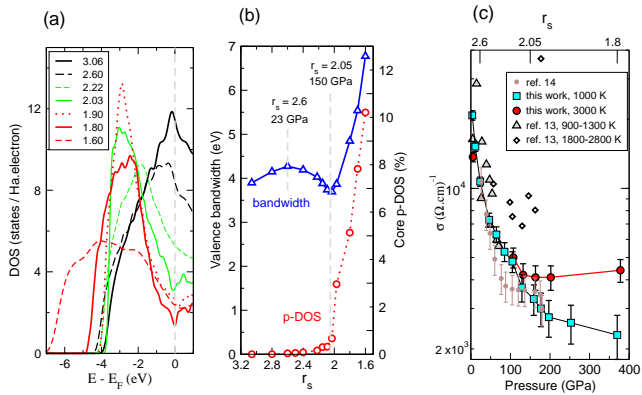}
  \caption{\label{fig_electronic} Electronic properties of dense molten Li. 
 (a) Electronic density of states, DOS. 
The corresponding $r_s$ values for each density are listed in the legend.  
(b) Valence electron bandwidth (blue triangles) and integrated DOS of core electrons with $p$ angular momentum character, p-DOS (red circles). (c) Calculated dc conductivity, $\sigma$, along the 1000~K and 3000~K isotherms. Error bars are statistical standard deviations. Comparison is made with experimental measurements\cite{11,12}. The data from \cite{11} are divided into two sets, representing two temperature intervals.  
Only qualitative comparison is meaningful with \cite{11} as a model equation of state has been used there to obtain Li density, $T$, and $\sigma$. }
\end{figure}

We now examine the changes in the electronic properties that are likely to drive the structural transitions. 
 With increasing density, the DOS at the Fermi level gradually decreases ( Fig.~\ref{fig_electronic}(a)). A similar effect, interpreted as a Peierls symmetry breaking, was observed in liquid Na \cite{4}, except that its strength in Li
is stronger, which can be understood in terms of an increasing hardness of the effective repulsive potential; it is similar, for example, to the decreased amplitude of the distortion in crystalline Sb in comparison to crystalline As \cite{10}. 
To interpret these changes and their likely consequences, we first look at the valence electron bandwidth (Fig.~\ref{fig_electronic}(b)). It increases with increasing density, es expeced upon densification. This tendency remains up to $P \sim 23$~GPa ($r_s = 2.6$), which matches well the pressure range over which the liquid becomes more compact. Above this pressure, the bandwidth begins to decrease due to a development of partial ($p$-character) bonding, which lowers the band-structure energy. The DOS develops a marked peak well below the Fermi level. This is also the range over which the coordination in the liquid begins to decrease. Finally, starting at $P \sim 150$ GPa ($r_s < 2.05$), the valence band begins to broaden again, but the DOS at the Fermi level does not decrease much further. This density range coincides with the conditions under which the tetrahedral coordination develops. 

\begin{figure}[t!]
  \hspace*{-5mm}
  \includegraphics[height=0.26\textheight,clip]{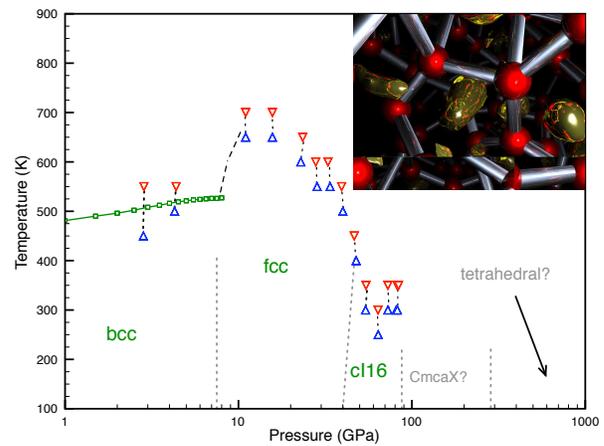}
  \caption{\label{fig_melt} Melting curve of Li under pressure. Available experimental data \cite{29,30} are shown in green. Uptriangle and downtriangle pairs indicate FPMD simulations of solid and liquids phases, respectively, on the same isochore, which bracket the melting temperature. The dashed line connecting the highest-pressure experimental data to the theoretical points near 10~GPa is only a guide to the eye. The vertical dotted lines indicate ambient temperature phase boundaries between solid phases. Only the high temperature phase boundary between $fcc$ and $cI16$ has been estimated based on their respective melting temperatures. For pressures above 90~GPa, previous theoretical studies have proposed crystals with the $Cmca$ \cite{1} and $Cmca24$ \cite{8} symmetries; hence, we have indicated this phase as “$CmcaX$”. The inset shows the tetrahedral structure, which develops in the liquid above 150~GPa. The picture is from a liquid configuration taken at $r_s=1.8$ and $T=1000$~K. The ions are shown as red balls, each connected to its nearest neighbours. Isosurfaces of the valence charge density are shown in gold; they illustrate the localization of valence electrons in the interstitial regions.  }
\end{figure}

%
The consideration of valence electrons alone is clearly insufficient to understand the modifications occurring at $P > 150$ GPa, which can be explained by examining the core electrons. 
Indeed, for $r_s < 2.05$ ($P > 150$~GPa) the $p$ angular momentum character of the core electrons increases rapidly (Fig.~\ref{fig_electronic}(b)), indicating core-core overlap.
At the same time, the volume available to the valence electrons, now squeezed into interstitial regions, decreases linearly with increasing density and hence the rapid increase of the valence bandwidth.
The resulting ''anti-diamond'' structure (inset in Fig.~\ref{fig_melt}) is with ions forming a tetrahedral network, but the valence electrons occupying preferentially the voids of the diamond structure instead of being located between adjacent ions.

The electrical conductivity ($\sigma$) of compressed liquid Li exhibits exactly the opposite $P$ and $T$ dependence usual for metals. Our computations \cite{el_str} show that along the 1000~K isotherm it has nearly a 10-fold drop between 3 and 150 GPa. When heating the liquid to 3000~K near and above 250~GPa, there is about twofold increase in $\sigma$, but this increase vanishes at lower $P$. These results are compared with measurements \cite{11,12} for which we now provide an explanation that is very different from what was originally proposed. 

\begin{figure}[t!]
  \hspace*{-5mm}
  \includegraphics[height=0.25\textheight,clip]{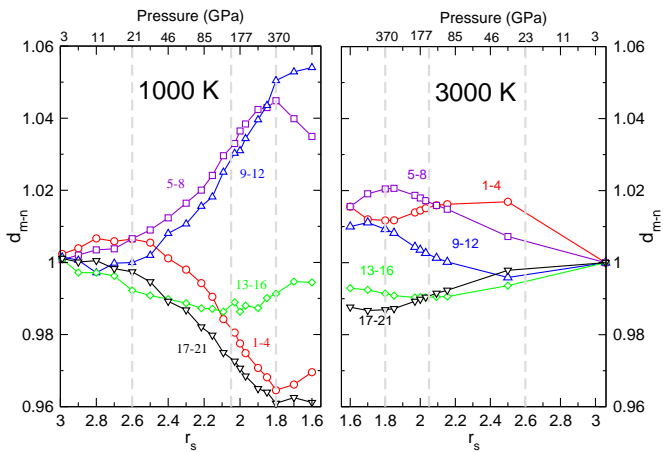}
  \caption{\label{fig_high_temp} Evolution of density-rescaled average interatomic distances with pressure and density at $T=1000$~K (left panel) and $T=3000$~K (right panel). As expected, increasing temperature leads to more homogeneous structures.  }
\end{figure}

First, the electron localization and decrease of DOS at the Fermi level lead to a significant drop in $\sigma$ with $P$. Second, the above-mentioned electronic changes are closely related to structural transitions from a higher to a lower-coordinated liquid. This trend is reversed when heating the liquid at a constant $V$; it reverts to a more homogeneous local order 
(see Fig.~\ref{fig_high_temp})
as favored by entropy. The result is an increase of $\sigma$ with $T$, as seen in our calculations above 150~GPa. This effect is of course countered by increased electron-ion scattering at high $T$. At $P \sim100$~GPa, the two effects cancel, and at lower $P$ where the structural changes are less significant, the scattering effects dominate. This explanation for the observed changes in $\sigma$ is different from what was previously suggested and we emphasize that the increase in conductivity observed at the highest-$P$ measurements by Fortov et al. \cite{11}) is not a pressure but a temperature effect. Despite the fact that the latter data were obtained using a model equation of state to estimate the density, resistivity, and $T$, the general trend of increased $\sigma$ with $T$ at sufficiently high $P$ is consistent with our findings.

Finally, we compute the melting curve (Fig.~\ref{fig_melt}) using a heat-until-melting approach, which provides an upper bound for the melting temperatures ($T_m$). The method was shown \cite{4} to be valid for Na for reasons that are likely to hold for Li. 
The melting curve has a steep negative slope between roughly 20 and 80~GPa, with $T_m$ dropping from (680 $\pm$ 25) K to (275 $\pm$ 25) K. The shape of the melting curve above $bcc$ is relatively flat, which is consistent with a gradual transformation in the liquid from $bcc$-like to $fcc$-like local order.
The $fcc$ solid is denser than $bcc$ and the melting slope above it initially increases, in accordance with the Clapeyron equation. The onset of symmetry breaking transitions lead to lowering of the liquid electronic band structure energy, its densification, 
and, hence, to the turnover of the melting curve above the $fcc$ phase. The maximum $T_m$ is near 20~GPa - exactly where we identified the onset of the changes towards lower coordination. The anomalous melting behavior persists until pressures where the solid also begins to undergo Peierls symmetry breaking transitions. 

%
We have only computed the melting curve in the pressure range where the crystalline structures are well established. Above 100~GPa, (or even down to 70~GPa), they remain unsettled; previous suggestions include $Cmca$ \cite{1} and $Cmca24$ \cite{8}. 
We propose that future investigations of low-$T$ crystalline phases of compressed Li focus on diamond-like tetrahedral structures. The appearance of such a phase, and especially its persistence at high $T$, is completely unexpected. While the modifications in Li are initially driven by Peierls-like distortions, at higher $P$ they are determined by core-core electron interactions and valence electron localization (resulting from core-valence interactions).
This behavior describes a distinct regime, likely present in other materials, where both valence and core electrons are responsible for chemical and physical properties. 
Such effects could have far reaching consequences in areas ranging from planetary modeling to the study of superconductivity under pressure \cite{13,14,15,16,17}. Another interesting aspect is the possible implications of these results for the properties of dense hydrogen. It has been discussed \cite{1,18} that similarities between Li and H could be used to predict high-pressure phases, including superconductivity \cite{19}, in metallic H.

%
%

Work supported by NSERC and CFI. Computational resources provided by ACEnet, IRM Dalhousie, Westgrid, and Sharcnet. I.T. acknowledges support by the Killam Trusts. We thank E. Schwegler, R. Redmer, A. Correa, M. Bastea, and V. E. Fortov for discussions.

%


\begin{thebibliography}{29}
\bibitem[{con()}]{convergence}
\bibinfo{note}{For selected $P$ and $T$, FPMD were carried out with 8 ${\mathbf
  k}$-points, with 256 atom cells and time steps up to 16 a.u. small (1 a.u. =
  0.02419 fs). The validity of our PAW potential at extreme density was tested
  on isolated Li$_5$ tetrahedral clusters by comparing with all-electron
  DFT-GGA (both PBE and B3LYP) calculations produced with the Gaussian quantum
  chemistry code \cite{25}, using a $-3/r$ ion potential. The quantity, which
  we compare is $dE/dr$, where $E$ and $r$ are the energy and the shortest
  interatomic distance in the cluster, respectively. The error in pressure for
  $2.7 < r < 3.8$ Bohr, which corresponds to the first peak of $g(r)$ for $1.7
  < r_s < 2.4$, is below 3\%.}
\bibitem[{el_()}]{el_str}
\bibinfo{note}{DOS and $\sigma$ were computed with ABINIT \cite{27}, using an
  all-electron PAW pseudopotential within the PBE-GGA. The reported results are
  from calculations with 32 {$\mathbf k$}-points, but tests were made with up
  to 1000 ${\mathbf k}$-points, and independently checked with VASP. The
  Kubo-Greenwood formula was used to obtain $\sigma$.}
\end{thebibliography}


\end{document}